\documentclass[12pt]{article}
\usepackage{graphicx}
\usepackage{amsmath}
\usepackage{amsfonts}
\usepackage{amssymb}
\usepackage{graphicx}
\newcommand{\be}{\begin{equation}}
\newcommand{\ee}{\end{equation}}
\newcommand{\bea}{\begin{eqnarray}}
\newcommand{\eea}{\end{eqnarray}}
\newcommand{\ba}{\begin{array}}
\newcommand{\ea}{\end{array}}
\newcommand{\fig}[1]{figure~\ref{fig:#1}}

\newcommand{\reff}[1]{ref.~\cite{#1}}
\newcommand{\refs}[1]{refs.~\cite{#1}}
\newcommand{\eq}[1]{eq.~(\ref{eq:#1})}
\newcommand{\eqn}[1]{(\ref{eq:#1})}
\newcommand{\eqs}[1]{eqs.~(\ref{eq:#1})}

\newcommand{\no}{\nonumber}
\newcommand{\nn}{\nonumber}

\newcommand{\ket}[1]{\vert {#1} \rangle}
\newcommand{\bra}[1]{\langle {#1}}
\newcommand{\la}{\langle}
\newcommand{\ra}{\rangle}

\newcommand{\cB}{{\cal B}}

\newcommand{\cN}{{\cal N}}
\newcommand{\cO}{{\cal O}}

\newcommand{\Heff}{{\cal H}_{ eff}}

\newcommand{\kpnn}{K \to \pi  \nu \bar\nu}
\newcommand{\kpll}{K \to \pi  \ell^+ \ell^-} 
\newcommand{\kppnn}{K^+ \to \pi^+  \nu \bar\nu}
\newcommand{\kppll}{K^+ \to \pi^+  \ell^+ \ell^-} 
\newcommand{\kspll}{K_S \to \pi^0  \ell^+ \ell^-} 
\newcommand{\klpll}{K_L \to \pi^0  \ell^+ \ell^-} 
\newcommand{\klpnn}{K_L \to \pi^0  \nu \bar\nu} 
\newcommand{\kppee}{K^+ \to \pi^+  e^+ e^-} 
\newcommand{\kspee}{K_S \to \pi^0  e^+ e^-} 

\def\npb#1#2#3{    {Nucl. Phys.}~B {\bf #1}, #3 (#2)}
\def\plb#1#2#3{    {Phys. Lett.}~B {\bf #1}, #3 (#2)}

\def\prd#1#2#3{    {Phys. Rev.}~D {\bf #1}, #3 (#2)}

\def\prl#1#2#3{    {Phys. Rev. Lett. }{\bf #1}, #3 (#2)}
\def\ptp#1#2#3{    {Prog. Theor. Phys. }{\bf #1}, #3 (#2)}

\def\rmp#1#2#3{    {Rev. Mod. Phys. }{\bf #1}, #3 (#2)}

\def\epjc#1#2#3{   {Eur. Phys. J.}~C {\bf #1}, #3 (#2)}
\def\jhep#1#2#3{   {JHEP  }{\bf #1}, #3 (#2)}

\def\cmp#1#2#3{    {Comm. Math. Phys.}~{\bf #1}, #3 (#2)}
\def\ibid#1#2#3{   {\em ibid.}~{\bf #1}, #3 (#2)}

\begin{document}

\begin{flushright}
June 2005 \\
hep-ph/yymmnn \\
\end{flushright}
\thispagestyle{empty}
\setcounter{page}{0}

\vskip   2 true cm 

\begin{center}
{\Large \textbf{Rare Kaon Decays on the Lattice}} \\ [20 pt]
\textsc{Gino Isidori},${}^{1}$ \textsc{Guido Martinelli},${}^{2}$ 
and \textsc{Paolo Turchetti},${}^{2}$  \\ [20 pt]
${}^{1}~$\textsl{INFN, Laboratori Nazionali di Frascati, I-00044 Frascati,
      Italy} \\ [5 pt]
${}^{2}~$\textsl{ Dipartimento di Fisica, Universit\'a di Roma
``La Sapienza'' and \\ INFN, Sezione di Roma, P.le A. Moro 2,
I-00185 Roma, Italy  } 

\vskip   2 true cm 

\textbf{Abstract}
\end{center}
\noindent
We show that  long distance contributions to the rare decays 
$\kpnn$ and $\kpll$ can be computed  using lattice QCD.  
The proposed approach requires well established methods, 
successfully applied in the calculations of  
electromagnetic and semileptonic form factors. 
The extra  power divergences, related to the use of weak  four-fermion operators,  
can be eliminated using only the symmetries of the lattice action 
without ambiguities or complicated non-perturbative subtractions. 
We demonstrate that this is true even when a lattice action with explicit 
chiral symmetry breaking is employed.  
Our study opens the possibility of reducing the present uncertainty  
in the theoretical predictions for these decays. 
\setcounter{footnote}{0}

\newpage

\section{Introduction}  
Rare decays mediated by flavour-changing neutral-currents (FCNC) 
are among the deepest probes to uncover the fundamental 
mechanism of quark flavour mixing. Within the Standard Model (SM),
these rare decays are strongly suppressed both by the GIM mechanism \cite{GIM} 
and by the hierarchy of the CKM matrix \cite{CKM}, and are often 
dominated by short-distance dynamics. As a result, FCNC processes 
are very sensitive to possible new sources of flavour mixing, 
even if these occur well above the electroweak scale. 
The sensitivity to physics beyond the SM of these rare processes
is closely related to the theoretical accuracy on which 
we are able to compute their amplitudes within the SM.
\par
Within the family of FCNC decays, long-distance effects are not always negligible
and, in most cases, they represent the dominant 
source of theoretical uncertainty. Long-distance contributions are typically 
relevant in: i) amplitudes where the GIM mechanism is only logarithmic;
ii) amplitudes where the power-like GIM suppression of the long-distance component 
is partially compensated by a large CKM coefficient. 
So far, the evaluation of these non-perturbative  
contributions has been performed by means of 
effective theories. These analytic tools require the 
introduction of additional parameters, 
the knowledge of which constitutes a source 
of sizable theoretical uncertainty.
\par
In this paper we show that for a class of very interesting processes, 
such as  $\kppnn$ and $\kpll$,  it is possible in principle to compute 
non-perturbatively  the long-distance contribution to the physical amplitudes 
on the lattice.  
The physical information is encoded in the following $T$-products:
\bea  
{\cal T}_{Q,J}^\mu(q^2) = \cN_V
\int\, d^4x \, \int\,  d^4y  \, e^{-i \,q \cdot  y}\, 
\la\pi \vert T[ Q(x) \,J^\mu(y)]\vert K\ra 
\, , \label{eq:T1}
 \eea 
where $Q$ denotes a generic four-quark operator of the effective weak Hamiltonian, 
$J^\mu$ is either the electromagnetic or the weak neutral current, and $\cN_V$ is an 
appropriate volume factor.
If the invariant mass of the lepton pair ($q^2$)  
is smaller than any physical hadronic  threshold, 
the calculation proceeds as in the case of semileptonic
form factors (see e.g.~\refs{kl3nostro,bdecays}), 
and one obtains directly the relevant amplitude. 
When instead  the leptonic invariant mass exceeds the pion threshold, 
the final state interaction induces problems similar to those encountered with non-leptonic 
kaon decays~\cite{ll,lmst}. However, the knowledge of the amplitude 
for $q^2 < m_\pi^2$ is sufficient to determine the leading 
unknown effective couplings of these amplitudes within the framework of
chiral perturbation theory (CHPT)~\cite{EPR}--\cite{IMS}. Therefore, 
the combination of lattice calculations and CHPT should allow 
to reach an unprecedented level of precision for these rare decays.
\par  
When using a  lattice action with explicit chiral symmetry breaking,  
such as Wilson, Clover or twisted mass fermions, 
further problems arise because of additional ultraviolet (power) divergences which may appear 
in the operator matrix elements or in the relevant $T$-products.\footnote{~Alternative  
formulations which guarantee chiral symmetry 
in the physical matrix elements, such as  overlap fermions,  do not have this problem~\cite{Chiu:2003bv}. 
However, these formulations are not mature yet to be used for unquenched 
calculations of these complicated matrix elements, for quark masses 
close to the physical values.} 
We show that for the electromagnetic current,  
gauge invariance prevents the appearance of these  
divergences even if the most popular lattice actions
are used. Consequently, when $J^\mu$ is the electromagnetic current, 
the $T$-products in \eq{T1}
are finite provided that a renormalized weak effective Hamiltonian is used. 
The situation is slightly more 
complicated when $J^\mu$ is the weak neutral current. 
In this case,   simple power counting, related to the 
behavior of the $T$-product at short distances,  
shows that both quadratic and linear divergences may appear. 
We show that the quadratic divergence, 
which is not a peculiarity of the lattice regularization,  is canceled by the GIM mechanism.
Concerning the linear divergence, which is 
present only if there is an explicit chiral symmetry breaking term in the lattice action,
we demonstrate that it can be avoided by using 
the maximally twisted mass fermion action~\cite{fr12}.
\par 
There is a further subtlety concerning the ambiguity in the  
renormalization of the effective weak Hamiltonian out of the chiral limit~\cite{varitesta}.  
In \reff{varitesta} it has been shown 
that this ambiguity does not affect the physical $K \to \pi\pi$ amplitudes, 
but is present in ``non-physical" matrix elements, such as $\la \pi \vert Q\vert K\ra$.    
This problem is present also in our case and implies an 
ambiguity in the $T$-products of \eq{T1}.  
By means of appropriate Ward Identities, we show that the physical amplitude, 
the extraction of which  requires a specific spectral analysis
discussed in the following,  is instead free of ambiguities. 
\par 
The  paper is organized as follows: in sect.~\ref{sect:general}
we recall the basic ingredients 
of radiative decays in the framework of the effective Hamiltonian approach.  
In sect.~\ref{sect:tprod} we describe the strategy for computing the relevant amplitudes 
from the Euclidean Green functions and  discuss the structure of the divergences in both cases: 
when they cancel because of gauge invariance and when it is necessary 
to get rid of them using GIM mechanism and twisted mass.
In sect.~\ref{sect:ambiguity} we  show how   
to extract  the physical amplitude in spite of the ambiguity of the renormalized  
effective  Hamiltonian. The results are summarized in the conclusions.

\section{Effective Hamiltonian for  $K\to \pi \ell^+ \ell^- (\nu \bar\nu)$ decays}
\label{sect:general}

The dimension-six effective Hamiltonian relevant to evaluate  
$s \to d  \ell^+ \ell^- (\nu \bar\nu)$ amplitudes at
next-to-leading order accuracy, renormalized at a scale 
$M_W \gg \mu > m_c$, can be written as 
\be
\Heff = \Heff^{|\Delta S|=1}\, +\, \Heff^{\rm FCNC}\, 
+\, \frac{G_F}{\sqrt{2}} \sum_{q=u,d,s,c}  Q^{\rm NC}_{q}\, +\,
\frac{G_F}{\sqrt{2}} \sum_{q=u,c\ q^\prime=d,s}  V_{ij} Q^{\rm CC}_{qq^\prime} 
\, +\,  \mbox{h.c.}~,
\label{eq:heff}
\ee
where $V_{ij}$ denote the elements of the CKM matrix,
\bea
\Heff^{|\Delta S|=1} &=& \frac{G_F}{\sqrt{2}} \, V_{us}^* V_{ud} 
    \left[   \, \sum_{i=1,2} C_i \left(Q_i^{u}-Q_i^{c}\right)
  +   \sum_{i=3\ldots 8} C_i Q_i  
  + \cO \left(\frac{V_{ts}^* V_{td}}{V_{us}^* V_{ud} } \right) \right], 
\eea
is the usual $|\Delta S|=1$ weak Hamiltonian, for which the Wilson coefficients
are known at the NLO \cite{basis}, and 
\bea
\Heff^{\rm FCNC} &=& \frac{G_F}{\sqrt{2}} \frac{\alpha}{ 2\pi \sin^2\theta_W} 
  \, V_{us}^* V_{ud} \left[  \,   \sum_{i=7V,7A,\nu} C_i Q_i 
  + \cO \left(\frac{V_{ts}^* V_{td}}{V_{us}^* V_{ud} } \right) \right]~.
\eea
Here
\bea
Q^{\rm CC}_{qq^\prime} &=& \bar q \gamma^\mu (1-\gamma_5) q^\prime  ~ \bar \nu
                     \gamma_\mu (1-\gamma_5) \ell \nn \no \\
Q^{\rm NC}_{q}  &=&  \bar q \gamma^\mu \left[ 2 \hat T (1-\gamma_5) 
 - 4 \hat Q \sin^2\theta_W  \right] q 
 \no \\  && \times \left[\, \bar\nu \gamma_\mu (1-\gamma_5) \nu \, - 
 \bar\ell \gamma_\mu (1-\gamma_5- 4 \sin^2\theta_W) \ell \,  \right]  
\label{eq:QZ} 
\eea
are the charged-current and neutral-current effective interactions 
obtained by the integration of the heavy $W$ and $Z$ fields, 
\bea
Q_{7V}  &=&  \overline{s} \gamma^{\mu}(1-\gamma_5) d \, \bar\ell
\gamma_{\mu} \ell~,  \label{eq:Q7v} \\
Q_{7A}  &=&  \, \overline{s} \gamma^{\mu} (1-\gamma_5) d \, \bar\ell
\gamma_{\mu} \gamma_5 \ell~,  \label{eq:Q7a} \\
Q_{\nu}  &=&  \, \overline{s} \gamma^{\mu} (1-\gamma_5) d \, \bar\nu
\gamma_{\mu} (1-\gamma_5) \nu~,  \label{eq:Qnn} 
\eea
are the leading FCNC operators, and 
\bea
Q_1^{q} &=& \bar s_\alpha \gamma^\mu (1-\gamma_5) q_\beta 
 ~ \bar q_\beta \gamma_\mu (1-\gamma_5) d_\alpha~, \\
Q_2^{q} &=& \bar s_\alpha \gamma^\mu (1-\gamma_5) q_\alpha 
 ~ \bar q_\beta \gamma_\mu (1-\gamma_5) d_\beta~
\eea
the leading four-quark operators. The four-quark operators
originated by penguin contractions are denoted by $Q_{1 \ldots 6}$,
whereas $Q_{7}$ and $Q_8$ correspond to magnetic and 
chromomagnetic operators, respectively  (see e.g.~\reff{basis}).
\par
Thanks to both the GIM mechanism and the unitarity of the CKM matrix, 
the contributions to the FCNC amplitudes can be unambiguously decomposed 
into two parts, the first one proportional to the CKM combination 
$V^*_{us} V_{ud}$,  the second one  proportional to   $V^*_{ts} V_{td}$. 
Since $|V^*_{ts} V_{td}| \ll |V^*_{us} V_{ud}|$, 
the contribution  proportional to $V^*_{ts} V_{td}$ is negligible but for 
 cases where it is enhanced by the large top-quark mass
(i.e.~for  amplitudes which exhibit a power-like GIM mechanism). 
In these cases, the amplitudes are completely dominated 
by short distances (top-quark loops) and can be evaluated 
in perturbation theory to an excellent degree of approximation. 
In this paper instead we are interested only in the long-distance 
components of the amplitudes, therefore we can safely work
in the limit  $V_{td}=0$.   
\par
We can seemingly neglect the 
matrix elements of $Q_{1 \ldots 6}$ and $Q_8$ in the evaluation of 
 $K\to \pi \ell^+ \ell^-\, (\nu \bar\nu)$ amplitudes:
these matrix elements vanish at the tree level and 
the corresponding Wilson coefficients are substantially 
smaller than  those of $Q_{12}^{u,c}$.
In this approximation, we only have to consider the 
contributions of the leading FCNC operators 
in \eqs{Q7v}--\eqn{Qnn} and the non-trivial contractions of $Q^{u,c}_{1,2}$ with 
the electromagnetic current and the currents 
defined by $Q^{\rm CC}_{qq^\prime}$ and $Q^{\rm NC}_{q}$.
The $K\to \pi$ matrix elements of  the FCNC operators 
in \eqs{Q7v}--\eqn{Qnn} can be extracted from data on the leading 
$K_{\ell 3}$ modes using isospin symmetry \cite{MP}, 
or even computed directly on the lattice, with high accuracy,
as recently shown in \cite{kl3nostro}.\footnote{~In principle, in the 
$\kpll$ case one should also consider the tree-level matrix element of the 
magnetic operator $Q_{7} = m_s \overline{s} \sigma^{\mu\nu}(1-\gamma_5) d \, F_{\mu\nu}$,
which cannot be directly extracted from $K_{\ell 3}$ data. 
However, within the SM the smallness of the corresponding 
Wilson coefficient makes 
this contribution negligible for practical purposes.
This  matrix element 
can  be computed on the lattice with standard techniques,
as shown in \cite{SPQR_mag}.}
Concerning the contractions of $Q^{u,c}_{1,2}$, those with a 
charged current receive very small non-perturbative contributions
(estimated to be below $1\%$ at the amplitude level in the $\kppnn$ case 
and even smaller in all the other channels), which 
can be reliably estimated within CHPT \cite{IMS,BI}.
Thus the main problem are the contractions of $Q^{u,c}_{1,2}$
with a neutral current, as outlined in \eq{T1}.
\par
So far, this problem has been addressed with the following 
two-step procedure: i) integrating out the charm as dynamical degree 
of freedom; ii) constructing the chiral realization of the corresponding 
effective Hamiltonian with light quarks only. This procedure suffers from 
two sources of theoretical errors: slow convergence of perturbation 
theory because of the low renormalization scale 
of the effective Hamiltonian ($\mu <m_c$);  uncertainties associated 
to the new low-energy couplings appearing in the effective theory.
Both these sources of uncertainties are naturally reduced in the lattice 
approach, where the effective Hamiltonian is renormalized above the 
charm scale and the $T$-products are evaluated in full QCD.
\par 
We now discuss separately electromagnetic and neutrino amplitudes 
in more detail.
\subsection{$\kpll$}
The main non-perturbative correlators relevant for these decays are those with the
electromagnetic current. In particular, the relevant $T$-product in 
Minkowski space is \cite{EPR,DEIP}
\bea
   \left({\cal T}^j_{i} \right)^{\mu}_{\rm em}(q^2)  
&=&  -i \int\,  d^4 x  \, e^{-i \,q \cdot  x}\,
   \bra{ \pi^j (p) }\vert  T \left\{ J^\mu_{\rm em}(x) 
   \left[ Q_{i}^{u}(0)-Q_{i}^{c}(0) \right] \right\} 
\ket{ K^j (k) }~, \label{eq:Tem}\\
 J^\mu_{\rm em} &=& \frac{2}{3} \sum_{q=u,c} \bar q \gamma^\mu q 
- \frac{1}{3} \sum_{q=d,s} \bar q \gamma^\mu q 
\eea
for $i=1,2$ and $j=+,0$. Thanks to gauge invariance we can write 
\be
\left({\cal T}^j_{i}\right)^{\mu}_{\rm em}(q^2) = 
\frac{w^j_i (q^2)}{(4\pi)^2} \left[q^2(k+p)^\mu -(m_k^2-m_\pi^2)q^\mu \right]~.
\label{eq:Tij}
\ee
The normalization of \eqn{Tij}
is such that the $O(1)$ scale-independent 
low-energy couplings $a_{+,0}$ defined in \cite{DEIP} 
can be expressed as   
\be
 a_j = \frac{1}{\sqrt{2}} V_{us}^* V_{ud} 
\left[ C_1 w^j_1(0) + C_2 w^j_2(0) + \frac{2 N_j }{\sin^2 \theta_W} f_+(0) C_{7V} \right]~.
\ee
where $f_+$ is the $K\to\pi$ vector form factor and 
$\{N_+,N_0\}=\{1,2^{-1/2}\}$~\cite{kl3nostro}.
To a good approximation, the decay rates of the CP-conserving
transitions $\kppll$ and $\kspll$ are proportional to the square of these 
effective couplings \cite{DEIP}:
\be
\cB(\kppee) \approx 6.6~ a_+^2 \times 10^{-7}~, \qquad 
\cB(\kspee) \approx 10.4~ a_0^2 \times 10^{-9}~.
\ee 
At present, we are not able to predict $a_{+,0}$ with sufficient accuracy: we 
simply fit their $\cO(1)$ values from the measured rates of the corresponding  decay modes
(an updated numerical analysis can be found in \cite{klp0ee}).
Being completely dominated by long distance contributions, 
these two CP-conserving processes would provide 
an excellent testing ground for the lattice technique.
\par 
On the other hand, the calculation of $a_{0}$ from first principles 
would have a very interesting phenomenological application
in the $\klpll$ case, which proceeds via a CP-violating amplitude: 
the calculation of $a_{0}$  would allow to determine 
in a model-independent way the sign of the interference 
between the (long-distance) indirect-CP-violating 
component of the amplitude and the interesting (short-distance) 
direct-CP-violating term  \cite{klp0ee}. This result would allow 
to perform a very precise test of direct-CP-violation in the 
kaon sector.

\subsection{$\kpnn$}
The power-like GIM mechanism of the leading electroweak amplitude,
implies a severe suppression of long-distance effects in these modes. 
In the CP-violating channel,  $\klpnn$, long-distance 
contributions are negligible well below the $1\%$ level \cite{BI}. 
However, this is not the case for the charged channel, $\kppnn$, 
where the suppression of long-distance effects is partially compensated 
by a large CKM coefficient. The $T$-product which determines the 
size of non-perturbative effects in this mode is \cite{BB_old,Falk}
\be
\left({\cal T}^+_{i} \right)^{\mu}_{Z} (q^2)  =  -i \int\,  d^4 x  \, e^{-i \,q \cdot  x}\,
   \bra{ \pi^+ (p) }\vert  T \left\{ J^\mu_{Z}(x) 
   \left[ Q_{i}^{u}(0)-Q_{i}^{c}(0) \right] \right\} 
\ket{ K^+ (k) }~,
\ee
where $J^\mu_{Z}=\bar q \gamma^\mu ( 2 \hat T (1-\gamma_5) 
 - 4 \hat Q \sin^2\theta_W  ) q$ is the neutral current defined by 
$Q^{\rm NC}_{q}$ in \eqn{QZ}.  Separating the electromagnetic 
component, we can write $\left({\cal T}^+_{i} \right)^{\mu}_{Z} = 
\left({\cal T}^+_{i} \right)^{\mu}_{L} 
- 4 \sin^2\theta_W \left({\cal T}^+_{i} \right)^{\mu}_{\rm em}$,
where 
\bea
\left({\cal T}^+_{i} \right)^{\mu}_{L} (q^2)  
 &=&  -i \int\,  d^4 x  \, e^{-i \,q \cdot  x}\,
   \bra{ \pi^+ (p) }\vert  T \left\{ J^\mu_{L}(x) 
   \left[ Q_{i}^{u}(0)-Q_{i}^{c}(0) \right] \right\} 
\ket{ K^+ (k) }~, \label{eq:Tnn} \\
 J^\mu_{L}  &=&  \sum_{q=u,c} \bar q \gamma^\mu (1-\gamma_5) q 
- \sum_{q=d,s} \bar q \gamma^\mu  (1-\gamma_5) q ~.
\eea
Contrary to $\left({\cal T}^+_{i} \right)^{\mu}_{\rm em}$,
the structure of  $\left({\cal T}^+_{i} \right)^{\mu}_{L}$ 
is not protected by gauge invariance and we can decompose it as 
\be
\left({\cal T}^+_{i}\right)^{\mu}_{\rm L}(q^2) = 
\frac{m_K^2 }{ \pi^2 } \left[ z^+_i (q^2) (k+p)^\mu  + \cO(q^\mu) \right]~,
\label{eq:TLdec}
\ee
where the normalization is such that the $z^+_i(q^2)$ are expected to be $\cO(1)$ \cite{IMS}.
The value of these form factors at $q^2=0$
is sufficient to control the long-distance contributions to the $\kppnn$ 
amplitude down to the 1\% level of precision~\cite{IMS}. 
\par
Charm and, more generally, long-distance contributions
to the $\kppnn$ amplitude, are usually parametrized in terms 
of a scale-independent coefficient $P_{c}$ \cite{BB_old}. According to 
the decomposition \eqn{TLdec}, this 
can be be written as 
\be 
P_{c}  =   \frac{1}{|V_{us}|^4} \left\{
\frac{m_K^2}{M_W^2} \left[ C_1 z^+_1(0) + C_2 z^+_2(0) \right] 
+ f_+(0) C_{\nu} \right\}~.  \label{eq:Pc}
\ee
The coefficient $P_c$ expresses the relative weight of the subleading
terms relative to the top-quark 
amplitude, which is the   leading contribution and is precisely determined  in perturbation  theory~\cite{BB_old}. 
As can be noted, the non-perturbative 
parameters $z^+_i(0)$ appear in \eqn{Pc} multiplied by a very small coefficient:
$m_K^2/M_W^2/|V_{us}|^4 \approx 0.015$.
Thus even a determination of these matrix elements 
at the 30--50\% level from lattice QCD would be sufficient to 
reduce the overall error on the $\kppnn$ rate around or below 
the 1--2\% level.

\section{$T$-products at short-distances  on the lattice}
\label{sect:tprod}
In this section we discuss the properties of the Euclidean Green functions 
necessary to extract the physical amplitudes defined in \eqs{Tem} and \eqn{Tnn}
in a numerical simulation. 
Since the ultraviolet behavior is quite different in the two cases, 
we discuss them separately,  starting from the $T$-product which involves 
the electromagnetic current. In both cases, we assume that the operators 
of the effective weak Hamiltonian have been renormalized, 
namely that all their physical matrix elements are finite as the lattice 
spacing goes to zero ($a \to 0$). 
The renormalization of the effective Hamiltonian is discussed  in the next section.
\par
The starting point to extract the physical matrix elements 
is the following Euclidean Green function 
\bea &\,&
  \left({\cal T}_i\right)^{\mu}_{\rm X}(q^2,t_\pi,t_K)  = \int\,  d^4x  \, 
   \bra{ \Phi_\pi (t_\pi,\vec p) }    J^\mu_X(0) 
   \left[ Q_{i}^{u}(x)-Q_{i}^{c}(x) \right] 
\Phi^\dagger_K (t_K,\vec k) \ra ~, \nonumber \\
&\,&    t_\pi >0 \, , \quad t_K <0 \, , \label{eq:ETem}
\eea
where the source (sink) for creating (annihilating) the pseudoscalar 
mesons at fixed space momentum are defined as
\be \Phi_i(t_i,\vec q_i) = \int d^3z \, e^{- i \,\vec q_i \cdot \vec  z}\,
\Phi_i(t_i,\vec z)\, , 
\ee
and $\Phi_i(t_i,\vec z)$ is a suitable local operator with the quantum numbers  
of the pion or kaon, respectively. Note that, in order to simplify the notation and the comparison 
between continuum and lattice formulae, we use the symbol of integral also 
to indicate sums over the lattice sites. 
\par
If not for the presence of the weak four-fermion operator, the calculation would proceed 
as for the standard weak and electromagnetic form factors, by studying the behavior 
of the Green functions at large $t_\pi$ and $|t_K|$~\cite{kl3nostro}. 
This would give the form factors computed at momentum 
transfer $\vec q= \vec k -\vec p$ and with energy transfer $q_0= E_K -E_\pi$.    
Since $Q_i$ is  summed over the whole lattice volume and hence it carries zero momentum,  
this general strategy remains valid also for the Green function in \eq{ETem}.  As explained  in the previous section, in order to extract the relevant low energy couplings, we are interested only to study the correlation function for $q^2 < m_\pi^2$. In this range  no rescattering of  intermediate states  is possible and thus we do not have problems in relating the Minkowskian  $T$ product to the Euclidean one. 

\begin{figure}[t]
\begin{center}
\includegraphics[height=5.5cm]{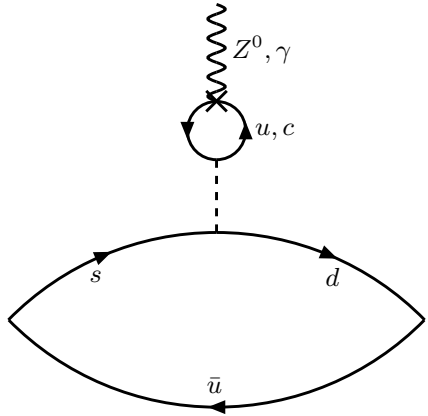}
\end{center}
\caption{\it One-loop topology which can originate power-like singularities
to the Green function \eqn{ETem} for $x\to 0$. The dotted line 
denotes the generic insertion of $Q^{u,c}_{i}$, with possible Fierz
re-arrangements. }
\label{fig:bolla1}
\end{figure}

 The additional problem which arises in this case 
is the possibility that the Green function itself diverges 
because of the short distance behavior when $x \to 0$. 
By dimensional arguments, this divergence can  at most be quadratic. At fixed lattice 
spacing $a$, this would imply potential contributions to the Green function of  $\cO(1/a^2)$.   
Fortunately this never happens, since the strongest divergence associated 
to the diagram in \fig{bolla1} is independent of the quark masses
and is canceled by the GIM mechanism. 
However, this cancellation does not guarantee the absence of 
linear divergences, which are naturally present when using 
lattice actions which break explicitly chiral invariance.

\subsection{The electromagnetic current}
\label{sec:lemc}
\begin{figure}[t]
\begin{center}
\includegraphics[height=4.0cm]{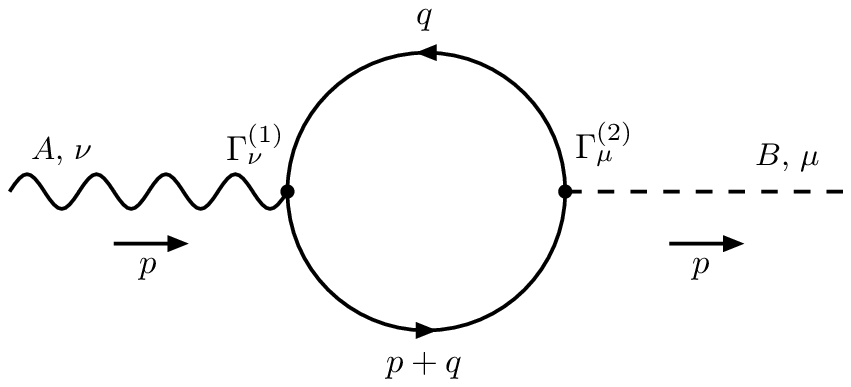}
\end{center}
\caption{\it Subdiagram of \fig{bolla1} associated to the $x\to 0$ singularity.}
\label{fig:bolla2}
\end{figure}

Even if the chirality of the fermion action is explicitly broken, we are still able 
to define a conserved vector current on the lattice, which we can identify with the
electromagnetic one. For example, with Wilson fermions we have
\bea 
  \hat{J}_V^\mu = \frac{1}{2} \left[ \bar q(x + \mu) U^{\mu\dagger}(x) (r + \gamma^\mu) q(x) 
- \bar q(x) U^\mu (x) (r - \gamma^\mu) q(x+\mu)  
\right]~, 
\label{eq:Jlat_em}
\eea
where $U^\mu$ is the link variable.  
With a conserved current, gauge invariance is strong enough to protect the 
Green functions from the appearance of both quadratic and linear divergences. 
This remains true even when the Wick contractions correspond to a vacuum polarization diagram
of the type in \fig{bolla1}, where only one of the two currents is the lattice conserved one,
and the other is a local vector current originating from the weak four-fermion operator.  
We have verified this argument by an explicit perturbative calculation using Wilson, 
Clover and twisted mass fermions. Since the results of this calculation (more precisely of 
the subdiagram in \fig{bolla2}) could be useful for other applications, 
we give them below for the Wilson and Clover cases.

\par
\medskip
The amplitude we have considered is 
\bea  
\Pi_{\mu\nu}(p) &=& \int \,\!\frac{d\,\!^4q}{(2\pi)^4}\, \mathrm{Tr} 
\left[ \Gamma^{(1)}_\nu(q;p+q)\Delta(q) \gamma_\mu \Delta(q+p) \right] \nonumber \\ 
&=& \frac{8}{{(4\,\pi)}^2}(\delta_{\mu\nu}p^2 - p_\mu p_\nu)
\left\{ \mathcal{I}(p^2 a^2, m^2 a^2) + L \right\}~, 
\label{eq:vittorio}\eea 
where $\Gamma^{(1)}_\nu(q;p+q)$ is the vertex derived from
\eq{Jlat_em} and $\Delta(q)$ the fermion propagator~\cite{giappo}.
Both in Wilson and Clover cases we can identify a universal infrared 
term, given by 
\bea
\mathcal{I}(p^2 a^2, m^2 a^2) = 
\int_0^1 d x \ x(1-x) \log \left[ m^2 a^2 + p^2 a^2 x(1-x)  \right] \, ,
\eea
while the finite constant $L$ depends from the details 
of the regularization. In the Wilson case we find 
\bea 
\label{eqn:Lperm}
&&  L_W = - \frac{1}{6} \log \left( m^2 a^2\right) +  (1 - \delta_{\rho\sigma}) 
\int_{-\pi}^{\pi} \! \frac{d^4 q}{2\,\pi^2} \  
\frac{-\frac{1}{2}\cos q_\rho \sin^2 q_\sigma 
- \frac{1}{3} \cos q_\rho \cos 2\,q_\sigma}{\Delta^2_F(q;m)} +  \nonumber \\
&& \qquad\qquad  + \frac{r^2\left( \frac{1}{3}\cos q_\rho 
\cos q_\sigma \sum_\tau \sin^2 \frac{q_\tau}{2} - \frac{1}{6}\cos q_\sigma \sin^2 q_\rho 
- \frac{1}{3} \cos q_\rho \sin^2 q_\sigma\right)}{\Delta^2_F(q;m)}\nonumber 
\eea
while in the Clover case
\bea 
L_{\rm Cl} = L_{\rm W} + \frac {r^2}{2\,\pi^2} (1 - \delta_{\rho\sigma})  
\int_{-\pi}^{\pi} \! d^4 q \, \frac{\frac{1}{2} \, 
\sin^2 q_\rho - \cos q_\sigma \sum_\tau \sin^2 
\left( \frac{q_\tau}{2} \right)}{\Delta_F^2 (q)}  \, .
\eea

\noindent 
Note that in both cases the absence of power divergences holds independently
from the GIM mechanism. Using the above results we could then match the lattice 
calculation with the continuum one even in an effective theory where the 
charm quark is integrated out. The comparison of the results obtained 
with or without dynamical charm quarks would provide a useful 
insight about the validity of the standard effective theory obtained by 
renormalizing $\Heff$ below the charm mass.
On the other hand, when the calculation is performed with a dynamical charm, the logarithmic divergence (and even the finite coefficient) in eq.~(\ref{eq:vittorio}) is cancelled by  the GIM mechanism.  For this reason no matching lattice to continuum is needed in this case.
\par
\medskip
Beside the possible singularities for $x\to 0$,
further divergences may arise from contact terms  
of $Q_i$ with the external sources, 
namely for $x \to x_\pi$ or $x \to x_K$. 
However, it is easy to show that these 
contact terms do not contribute to the physical 
amplitudes. Let us consider the Minkowski    $T$-product
\be
 \left({\cal T}_i^{\mu}\right)_{X}(q^2,t_\pi,t_K)  =  -i\,  \int\,  d^4x  \, 
   \la 0\vert T\left\{\Phi_\pi (t_\pi,\vec p)     J_X^\mu(0) 
   \left[ Q_{i}^{u}(x)-Q_{i}^{c}(x) \right] 
\Phi_K (t_K,\vec k) \right\}\vert 0\ra ~, \label{eq:MTem}
\ee
corresponding to the Euclidean Green function of eq.~(\ref{eq:ETem}).
The contact terms are proportional to the following pole terms: 
 $(p^2-m_\pi^2)^{-1}(k^2-m_\pi^2)^{-1}$ or  $(p^2-m_K^2)^{-1}(k^2-m_K^2)^{-1}$, 
while the on-shell  amplitudes are obtained form the coefficient of 
$(p^2-m_\pi^2)^{-1} (k^2-m_K^2)^{-1}$. 
As we shall discuss in more detail in the next section, these different pole structures 
in the Minkowski space correspond to a different 
$t_{\pi} \to \infty$ and $t_{K} \to - \infty$ 
 behavior in the  Euclidean case. As a result, we can eliminate the 
contact terms by an appropriate spectral analysis 
of the Green function computed in the numerical simulation.

\subsection{The axial current}
\label{sec:lax}
With the axial current appearing in the $T$-product \eqn{Tnn}, which is 
relevant for $K\to\pi\nu\bar\nu$ decays, 
we cannot invoke gauge invariance: it remains true 
that the quadratic divergence is canceled by GIM, but we 
must face the problem of the linear one.  
With power divergences, any subtraction procedure, though non-perturbative,   
would produce an irreducible (and thus unacceptable) ambiguity in the final result. 
This implies that the linear divergence can only be an artifact of 
the regularization procedure. This divergence is 
indeed absent in regularizations which preserve chirality.  
\par
With Wilson  fermions   the explicit breaking of chiral invariance 
leads to the appearence of such linear divergence.
Since this problem   is associated only to the contact 
term of the integrand (\ref{eq:ETem}) for $x\to0$, 
 we can  in principle obtain a finite subtracted $T$-product, with    the correct chiral behaviour 
of the Green function,  by an integration which 
avoids the region close to $x=0$.\footnote{ We thank Massimo Testa for discussions on this point.} 
Otherwise, one could introduce an appropriate set of counterterms 
and fix their values by  imposing an appropriate set of Ward identities,  
to recover the  correct chiral behaviour. However, both  
these procedures are technically very complicated to be implemented.
\par
A much simpler and techincally feasible solution is 
obtained by means of maximally twisted mass terms \cite{fr12}. 
In this case, the additional symmetries of the action imply that the amplitude we
are interested in is even in the Wilson parameter ($r$). This, in turn, 
implies the absence of the linear divergence which can only 
be odd in $r$, being associated to the breaking of chirality.
We have verified this statement by an explicit perturbative calculation
at the one loop level. As expected, the structure of the divergent terms 
is the same as in the continuum and the result is free from ambiguities. 
The discussion of the axial current can be repeated for a ``non-conserved" vector current, 
such as the lattice local electromagnetic current, or the vector component of the weak 
left-handed current in \eq{Tnn}. 
\medskip
\par   
At this point we wish to comment about the possibility 
to determine the physical $K\to\pi\pi$ amplitudes by 
using information about the following $T$-product 
\bea   
\int\, d^4x  \, e^{-i \,p_1 \cdot  x}\, 
\la\pi(p_2) \vert T[ \Heff(0) A^\mu(x)]\vert K(q_1)\ra 
\, , \label{eq:Tgil}
\eea    
where the axial current $A^\mu$ has the quantum numbers 
of the pion, but the kinematical configuration 
does not correspond to an on-shell pion field \cite{Gilberto}. 
The Wick contractions for this $T$-product are similar to those 
considered for $K \to \pi \nu \bar \nu$. In particular, the quadratic 
divergence generated when  $x \to 0$ is present also in this case. 
However, the situation is worse than the case discussed in this work, 
since there is no GIM mechanism to cancel the leading singularity. 
Of course we can define a renormalized $T^*$-product, 
but this would entail a finite ambiguity. 
The practical problems which need to be faced in order to 
avoid this ambiguity make this calculation very difficult
(if not practically impossible) with Wilson-type fermions. 
For this reason, we do not believe that a lattice study 
of this $T$-product can provide a useful tool to 
simplify the problem of determining $K \to \pi\pi$ amplitudes.

\section{$\Heff$ ambiguities}
\label{sect:ambiguity}
In this section we address the problems arising by the renormalization 
of the lattice operators of the effective weak Hamiltonian. We first note that only the 
parity-even or parity-odd terms of the operators contribute to the vector or axial-vector cases, 
respectively.  This observation is relevant since parity-even and parity-odd parts of the operators 
renormalize in a different way under regularizations which break chiral symmetry. 
On general grounds, whether chirality is broken or not, the mixing with operators 
of dimension five or six, 
in the presence of the GIM mechanism, does not introduce any ambiguity and the corresponding mixing 
coefficients can be computed in lattice perturbation theory. 
The problem arises from the mixing of the standard dimension-six operators with 
the scalar and pseudoscalar densities, which we now consider separately for 
the two cases.
\par
Schematically, we can write the renormalized operator as  
\bea
\label{eqn:espressionemescolamento}
\hat  Q^\pm  &=&Z^\pm(\mu a) \left[ Q^\pm    +  C_P\,  (m_c - m_u)(m_s - m_d) \,   
\bar s \gamma_5 d + C_S\,  (m_c - m_u)\,   \bar s d\right]   \, , 
\eea
where $Q^\pm =(Q_1\pm Q_2)/2 + \dots $ represents the ensemble of all the dimension six 
and five operators with  mixing coefficients computed in perturbation theory. 
By dimensional arguments,  
it follows that the coefficients $C_{S}$ and $C_{P}$  are power divergent in 
the limit $a\to 0$:
\be
 C_P \sim \frac{1}{a}~, \qquad C_S \sim \frac{1}{a^2}~.
\ee
Using suitable Ward identities (subtraction conditions), we can cancel the divergent 
parts of $C_P$ and $C_S$; however, this leaves an ambiguity in their finite values 
out of the chiral limit \cite{varitesta,giustiperoverlap}.  
For physical $K \to \pi\pi$ amplitudes this ambiguity turns out to be irrelevant:
the pseudoscalar density is proportional to the four divergence of the axial current
and its matrix element vanishes for the on-shell $K \to \pi\pi$ transition \cite{varitesta}.
We stress that this conclusion does not hold for the $K \to \pi$ case: 
the off-shell matrix element  $\la \pi \vert \bar s d\vert K\ra$ is different from zero
thus, in general, the $\la \pi \vert \hat Q \vert K\ra$ matrix element 
does suffer from this ambiguity.
\par 
Since we are interested in physical amplitudes,  we must be able to demonstrate 
that also in the case of radiative decays the 
matrix elements of scalar and pseudoscalar densities 
do not contribute to the on-shell amplitudes.
This can be done by means of suitable Ward identities and 
the spectral analysis of the relevant Euclidean 
Green functions. 
\par 
In the vector case we can use the following Ward identity
\bea
\label{eq:WIvett}
	 && \int \, d^4x\,  \left\{ \langle \Phi_\pi(x_\pi) 
\left[ \nabla_\mu {\hat V}_\mu^{sd}(x) +(m_d - m_s) \bar s d (x) \right]  
{\hat J}_V^\nu(y) \Phi^\dagger_K(x_K) 
\rangle \right\} ~=~  \nn \\
	&& \qquad\qquad ~=~  
- \langle \Phi_K(x_\pi){\hat J}_V^\nu(y)\Phi_K^\dagger(x_K)\rangle 
+ \langle \Phi_\pi(x_\pi) {\hat J}_V^\nu(y) \Phi_\pi^\dagger(x_K)  \rangle \, ,  
\eea
where the term between square bracket is the rotation of the lattice action, 
\bea
{\hat V}_\mu^{sd} &=& -\frac{1}{2} [\bar s (x) U_\mu (x) (r - \gamma_\mu) d(x+\mu) 
       - \bar s(x + \mu) U^\dagger_\mu (x) (r + \gamma_\mu)  d(x) ]\, ,
\eea
and the last two terms in \eqn{WIvett}
correspond to the rotation of the pion sink and the kaon source, respectively. 
The term with the four divergence of ${\hat V}_\mu^{sd}(x)$,  integrated over all space, 
vanishes. 
Thus on the left-hand side we are left with the term we are looking for,
up to overall factors, namely the contribution of the scalar density to the Euclidean 
Green function \eqn{ETem}, which enters when  the bare weak operators are replaced with 
the renormalized ones.  We need to show that this term does not 
contribute to the physical amplitude. 
\par
In the Minkowski space, the physical amplitude is identified by the 
coefficient of the physical pole for $p^2\to m_\pi^2$ and 
$k^2\to m_K^2$. In the Euclidean space, this corresponds to 
a well-defined dependence on $t_K$ and $t_\pi$ 
(for $t_{\pi} \to \infty$ and $t_{K} \to - \infty$), 
namely
\be 
  \frac{1}{(p^2-m_\pi^2) (k^2-m_K^2)} 
  \quad \leftrightarrow \quad 
   e^{-E_K |t_K|}\times  e^{-E_\pi t_\pi}~.
\label{eq:Tdphys}
\ee
The Ward identity  \eqn{WIvett} tell us that the contribution 
of the scalar density give rise to a different pole structure:
\bea 
\frac{1}{(p^2-m_\pi^2)  (k^2-m_\pi^2)} &\quad \leftrightarrow \quad  &
e^{-E_\pi|t_K|}\times   e^{-E_\pi t_\pi}  \nn \\
\frac{1}{(p^2-m_K^2)  (k^2-m_K^2)} &\quad \leftrightarrow \quad  &
e^{-E_K|t_K|}\times   e^{-E_K t_\pi} 
\eea
Thus the scalar density contribution can simply be eliminated by 
a study of the time dependence of the appropriate Green function. 
Incidentally, this procedure eliminates also the divergent contact 
terms mentioned at the end of the  section~\ref{sec:lemc}. 
\par
In the axial case, we have a similar situation, up to terms which vanish 
(linearly or quadratically in the lattice spacing) and inessential 
numerical factors. In particular, we can use the following Ward identity
\bea
 &&\int \, d^4x\,  \left\{ \langle \Phi_\pi(x_\pi) 
 \left[- \nabla_\mu Z_A \hat A_\mu^{sd}(x) +(m_d + m_s) \bar s\gamma_5 d (x)  
+ {\cal O}(a) \right]  J_A^\nu(y) \Phi^\dagger_K(x_K) \rangle \right\} = \nonumber \\
	&&=-  \langle \Sigma_K(x_\pi) J_A^\nu(y)\Phi_K^\dagger(x_K)\rangle 
+ \langle \Phi_\pi(x_\pi) J_A^\nu(y) \Sigma_\pi^\dagger(x_K)  \rangle \, ,  
\eea
where again  the term between square bracket is the rotation of the lattice action 
(for the explicit expressions of $Z_A$ and the weak renormalized axial current 
see~\cite{Bochicchio}) and $\Sigma_i$ is a scalar particle source.
This immediately  shows that  also the pseudoscalar density give rise 
to a time dependence different from the one in \eq{Tdphys} and thus 
does not contribute to the on-shell amplitude.

\section{Conclusions}
\label{sect:conc}

The potential of rare $K$ decays in performing precise 
tests of the SM and setting stringent bounds on physics 
beyond the SM depends, to a large extent, from our 
ability compute their amplitudes within the SM.
In this paper we have shown that for a class of very interesting processes, 
such as  $\kppnn$ and $\kpll$, the theoretical error associated 
to non-perturbative effects could be reduced by means of lattice calculations.
In particular, the numerical study of the 
Euclidean Green functions in \eq{ETem}, combined with CHPT, 
should allow to reach an unprecedented level of precision 
for these rare decays.
\par  
The main problem which needs to be addressed before starting 
a lattice calculation of these Euclidean Green functions
is the absence of power divergences in the 
extraction of the physical amplitudes. These may originate 
from contact terms between the weak four-fermion operators 
and the external fields ($\pi$, $K$ and the lepton current), 
or from the mixing of the four fermion operators 
with operators of lower dimensionality. 
In this paper we have shown that both 
these problems can be solved. 
\par
As demonstrated in section~\ref{sect:ambiguity}, 
the spectral analysis necessary 
to extract the physical amplitudes eliminates both 
the power divergences due to the operator mixing and the 
contact terms with the external $\pi$ and $K$ fields. 
The only remaining issue is then the ultraviolet behavior 
associated to the contact terms between the weak   
operators and the lepton current. This point is different 
for weak and electromagnetic currents.
\par 
In the electromagnetic case, relevant for $\kpll$ decays,
gauge invariance prevents the appearance 
of power divergences for all the popular Wilson-type actions.
The cancellation of power divergences 
is also independent of the GIM mechanism. We can thus 
match the lattice calculation with the continuum one also in an effective 
theory where the charm quark is integrated out. The perturbative 
expressions necessary for this matching at the one-loop level 
have been presented both for Wilson and Clover fermions. 
The situation is slightly more complicated for the weak 
(axial or vector) current, relevant for  $\kppnn$ decays, where 
we cannot invoke anymore gauge invariance. One can cancel 
power divergences also in this case with Wilson-type
fermions, but only using maximally twisted mass terms
and taking advantage of the GIM mechanism.  
\par 
In summary, our analysis shows that the numerical study of the 
Green functions relevant for $\kpll$ decays can be performed with 
any Wilson-type action, independently of the GIM mechanism. 
On the other hand, the study of $\kppnn$ decays on the lattice 
requires a more sophisticated action: with Wilson-type
fermions the only possibility is to use maximally twisted mass terms.
We believe that these results opens a new field of interesting  
physical applications to the lattice community.

\section*{Acknowledgments}
We warmly thank Vittorio Lubicz, Giancarlo Rossi, Silvano Simula 
and Massimo Testa for illuminating discussions. This work was supported 
in part by  the IHP-RTN program, EC contract No.~HPRN-CT-2002-00311 (EURIDICE).


\end{document}